\documentclass[12pt,eqsecnum,aps,preprintnumbers,superscriptaddress,prd,nofootinbib]{revtex4}

\usepackage[pdftex]{graphicx}
\graphicspath{{../pdf/}{../jpeg/}}
 \DeclareGraphicsExtensions{.pdf,.jpeg,.png}
 
\usepackage{amsmath}
\usepackage{enumerate}
\usepackage{amsfonts}
\usepackage{epsfig}

\newcommand{\be}{\begin{equation}}
\newcommand{\ee}{\end{equation}}
\newcommand{\bea}{\begin{eqnarray}}
\newcommand{\eea}{\end{eqnarray}}

\def\le{\left}
\def\ri{\right}
\def\[{\left [}
\def\]{\right ]}
\def\nein{N_{\rm Gravity}}
\def\nmax{N_{\rm Matter}}

\def\pure{{G}_{0}\!}
\def\BB{{G}_{\rm BB}\!}
\def\dA{\chi_2^{\rm BB}}
\def\df{\chi_1^{\rm BB}}

\def\dApure{\chi_2^{\rm pure}}
\def\dfpure{\chi_1^{\rm pure}}
\def\dAfpure{\chi_{1,2}^{\rm pure}}
\def\cut{\rm c}

\begin{document}

\title {Holographic Disorder Averaging}
\title {Disordered Holographic Systems}
\title {Holographic Disorder I -- Perturbative Backreaction}
\title {Disordered Holography}
\title {Ill-Condensed Holographic Systems I:\\ Rugged Black Brane}
\title {Dirty Holograms}
\title {Hot and Dirty Holograms}
\title {Hard Probes of Dirty Holograms}
\title {Giant Probes of Dirty Holograms}
\title {Unprotected Giant Probes of Dirty Holograms}
\title {Ill-Condensed Holographic Matter:\\ Perturbative Thermodynamics}
\title {Ill-Condensed Holographic Matter I:\\ Perturbative Thermodynamics}
\title{Gazing at Dirty Holograms I:\\ The Rise of Imperfections}
\title{Gazing at Dirty Holograms I:\\ Marginal Relevance of Imperfections}
\title{Hot and Dirty Holograms I:\\ Marginal Relevance of Imperfections}
\title{Hot and Dirty Holograms I:\\Functional Flows}
\title{Holography with Impurities I:\\Functional Flows}
\title{Dirty Holograms I:\\Functional Flows}
\title{Disordered Holographic Systems I:\\Don't Do Replica!}
\title{Disordered Holographic Systems I:\\Functional Renormalization}

\preprint{MIT-CTP-4211}
\preprint{SU-ITP-11/02}

\author{Allan Adams}
\affiliation{Center for Theoretical Physics, MIT, Cambridge, MA 02139}
\author{Sho Yaida}
\affiliation{Department of Physics, Stanford University, Stanford, CA 94305}

\begin{abstract}~\\
We study quenched disorder in strongly correlated systems via holography, focusing on the thermodynamic effects of mild electric disorder.
Disorder is introduced through a random potential which is assumed to self-average on macroscopic scales.
Studying the flow of this distribution with energy scale leads us to develop a holographic functional renormalization scheme.
We test this scheme by computing thermodynamic quantities and confirming that the Harris criterion for relevance, irrelevance or marginality of quenched disorder holds.
\end{abstract}

\maketitle
\newpage

\section{Imperfection}\label{introduction}
There exist no perfectly ordered materials in nature. Every crystal's formation involves impurities sneaking in and getting stuck, lattice vacancies remaining unfilled: disorder is inevitable. These microscopic imperfections often leave conspicuous imprints on materials' macroscopic properties. A well-known example is Anderson localization in systems of noninteracting quasiparticles~\cite{Anderson,4gangs,DisorderReviews} in which quenched disorder traps the quasiparticles, turning metal into insulating Fermi glass~\cite{Fermiglass}. Similarly, quenched disorder in certain frustrated spin systems leads to glassy phases at low temperature~\cite{Spinglass1,Spinglass2}.

In strongly correlated systems, however, our theoretical understanding of quenched disorder remains rather primitive. Is there a strongly correlated avatar of Anderson localization? What do Mott's law for direct current conductivity and its percolating picture in weakly correlated systems~\cite{Mott,ES,Percolation} morph into at strong coupling? Does many-body localization really happen~\cite{ManyBodyLocalization}? What does quenched disorder do to systems governed by quantum critical points?  Interesting theoretical questions abound. Meanwhile, from a pragmatic point of view, many technologically interesting systems, including the cuprate superconductors, are both strongly correlated and strongly disordered.  It is clearly worthwhile to investigate the effects of quenched disorder in strongly interacting systems.

To attack these challenging questions, we bring to bear holography, a powerful tool for studying the thermodynamic and transport properties of strongly correlated systems~\cite{MaldacenaOriginal,GKP,Witten,HolographyReviews}.    Our ultimate goal with this holographic approach is to find novel phases triggered by quenched disorder and to study transport properties within such phases.  The goal of the present paper is more modest: we merely point toward promising places to look for interesting phenomena in the holographic context and begin developing some of the tools needed to explore them.  
Specifically, we trace both the flow of dilute disorder deformation and its effect on themodynamic quantities, working perturbatively around clean fixed points.

As discussed below, self-averaging quenched disorder can be characterized by a distribution $P_{V}[W({\bf x})]$ over random functions $W({\bf x})$.  Importantly, as we change the energy scale, the entire functional runs. In contrast to traditional setups where we have only a few relevant running parameters to keep track of, we must now deal with an uncountable infinity of running couplings. To this end, we develop a {\it holographic  functional renormalization} scheme, which enable us to compute disorder-averaged thermodynamic quantities in holographic theories at finite temperature.

We check the validity of our scheme by applying it to confirm that the Harris criterion around the clean fixed point holds for quenched disorder characterized by a Gaussian distribution, at leading order in the strength of the disorder: the disorder is relevant when the scaling dimension of the strength of the disorder, $ \nu_{\rm dis} = d+1-2\Delta_{\cal{O}} $, is positive, irrelevant when $\nu_{\rm dis}$ is negative, and marginal when this scaling dimension vanishes. Here, $d$ is the spacetime dimension of the conformal field theory (CFT) and $\Delta_{\cal O}$ is the scaling dimension of the disordered operator. Whether marginal disorder is marginally relevant or irrelevant is a fascinating question we will revisit in a sequel~\cite{2ndAY} by utilizing the technology developed herein.

The organization of the paper is as follows.
In Sec.\ref{model}, we set up a holographic model with quenched disorder; our prototypical test case will be quenched electric disorder in CFTs dual to classical Einstein-Maxwell theory, but our results generalize straightforwardly.
In Sec.\ref{rugged}, we study the bulk response to quenched-disordered boundary conditions, including backreaction on the metric, working perturbatively in the strength of the disorder.
In Sec.\ref{peek}, we then explore how the disorder distribution evolves as we change the energy scale.
In Sec.\ref{fRG}, we propose a holographic functional renormalization scheme.
Armed with the scheme stipulated in Sec.\ref{fRG}, we compute leading quenched-disorder correction to grand potential in Sec.\ref{thermodynamics}. In particular, we verify that the Harris criterion described above holds.
We conclude in Sec.\ref{conclusion} with a view towards the scenery beyond the perturbative regime.

\section{A Model Holographic System with Quenched Disorder}\label{model}
For ease of presentation, we will henceforth focus on a strongly correlated CFT which is holographic to classical Einstein-Maxwell theory with action\footnote{The special case with $d=1+1$ should be treated with caution as the system would never enter hydrodynamic regime. Our holographic calculations are performed for integers $d\geq 2+1$ and then results are analytically continued to all real numbers $d>1+1$.}
\be\label{action}
S_{\rm bulk}=\frac{1}{16\pi G_N}\int d^{d+1}x\sqrt{-g}\le[R+\frac{d(d-1)}{L^2}\ri]-\frac{1}{4g_{d+1}^2}\int d^{d+1}x\sqrt{-g}F_{MN}F^{MN}.
\ee
Here, the dimensionless constants $\frac{L^{d-1}}{G_N}\equiv \nein^2$ and $\frac{L^{d-3}}{g_{d+1}^2}\equiv \nmax^2$ are determined by the parameters of the boundary CFT.  We take a large $\nein$ limit to ensure classicality of the bulk theory, while keeping $\frac{\nmax}{\nein}\sim 1$ to bring the role of gravitational  backreaction to the fore.
The resulting classical equations of motion are
\bea
\frac{1}{\sqrt{-g}}\partial_Q\le[\sqrt{-g}g^{QP}g^{MN}\le( \partial_P A_N-\partial_N A_P\ri)\ri]&=&0\\
{\rm and}\ \ \ \ \ R_{MN}-\frac{1}{2}R g_{MN}-\frac{d(d-1)}{2L^2}g_{MN}&=&\frac{8\pi G_N}{g_{d+1}^2}\le[F_{MP}F_{N}^{\ P}-\frac{1}{4}F_{PQ}F^{PQ}g_{MN}\ri]
\eea
where $F_{MN}\equiv \partial_M A_N-\partial_N A_M$.

Let us now sprinkle impurities into a clean strongly correlated CFT defined holographically as above.
We will focus on the effects of a random ``electric" potential $V({\bf x})=\int\frac{d^{d-1}{\bf k}}{(2\pi)^{d-1}}e^{i{\bf k}\cdot{\bf x}}V({\bf k})$ caused by quenched impurities in the system. In particular, the potential is time-independent: quenched impurities are, by definition, frozen on experimental time scales.
The action of the clean boundary CFT, $S_0$,  is thus deformed to
\be\label{QD}
S_V=S_0+\int dtd^{d-1}{\bf x} V({\bf x})J^t(t, {\bf x}).
\ee
Here $J^{\mu}$ is a conserved $U(1)$ current of the clean CFT which is dual to a bulk $U(1)$ gauge field, $A_{M}$.
Via the holographic dictionary, this electric disorder translates into disordered boundary conditions on the bulk $U(1)$ gauge field,
\be\label{UVBC}
\lim_{r\rightarrow\infty}A_M(r, t, {\bf x})=\delta_{tM}V({\bf x}).
\ee
Our choice of bulk coordinate system will be stipulated explicitly in the next section.

Finally, the quenched random potential $V({\bf x})$ is assumed to {\it self-average} on macroscopic scales.\footnote{Physically, this means that homogeneity is approximately restored as we average measurements over regions much larger than typical disorder length scales.  While the microscopic details of self-averaging  disorder wash out, its effects persist via its statistical properties, for example in the rounding of the sharp Drude peak in real metals.  To be sure, self-averaging is not universal, though it is very common. See, for example, Sec.III.A of~\cite{Spinglass1} for a detailed discussion of when and why such disorder-averaged quantities give extremely accurate estimates of observable quantities for a macroscopic sample with quenched disorder.}
Given self-averaging disorder, we can legitimately estimate densities of extensive quantities, for example the grand potential $\Omega$, as
\be
\le[\frac{\Omega}{V_{d-1}}\ri]_{\rm d.a.}\equiv\int{\cal D}W P_V[W] \le(\frac{\Omega_{W({\bf x})}}{V_{d-1}}\ri)
\ee
in the thermodynamic limit where the volume of the sample $V_{d-1}$ approaches infinity.
Here, $P_V[W({\bf x})]$ is the functional associated with $V({\bf x})$ satisfying
\be
\int{\cal D}W \, P_V[W]\,W({\bf x}_1) ...W({\bf x}_n)  \equiv \frac{1}{V_{d-1}}\int d^{d-1}{\bf x}_0 V({\bf x}_1+{\bf x}_0)...V({\bf x}_n+{\bf x}_0)
\ee
and $\Omega_{W({\bf x})}$ is the grand potential of the system with an electric potential $W({\bf x})$.

\subsection{Gaussian distribution and Harris criterion}

As a concrete example, let us consider the disorder dictated by a Gaussian distribution, randomly varying from site to site:\footnote{This corresponds to uncorrelated impurities with $\int{\cal D}W \, P_V[W]\,W({\bf x})W({\bf y})=f_{\rm dis}\delta({\bf x}-{\bf y})$.}
\be\label{distribution}
P_V[W({\bf x})]=N_{\sharp}e^{-\frac{1}{2f_{\rm dis}}\int d^{d-1}{\bf x}W({\bf x})^2}=N_{\sharp}e^{-\frac{1}{2f_{\rm dis}}\int \frac{d^{d-1}{\bf k}}{(2\pi)^{d-1}}W({\bf k})W(-{\bf k})}.
\ee
The normalization constant $N_{\sharp}$ ensures $\int{\cal D}W P_V[W] =1$ and the constant $f_{\rm dis}$ characterizes the strength of the quenched disorder. 
By dimensional analysis, $f_{\rm dis}$ is seen to have the scaling dimension $\nu_{\rm dis}=d+1-2\Delta_{J^t}$ near the clean fixed point, where $\Delta_{J^t}=d-1$ is the dimension of $J^t$ in the clean CFT.
Thus, we would expect that the quenched electric disorder becomes irrelevant (relevant) at long distance if $\Delta_{J^t}>\frac{d+1}{2}$ ($\Delta_{J^t}<\frac{d+1}{2}$), in other words, if $d>2+1$ ($d<2+1$).
This Harris criterion arises, for example, in disorder-averaged vacuum two-point correlation functions.\footnote{An analysis for ``classical" disorder has been carried out in~\cite{Japanese} by using the replica trick. Interestingly, those results can be obtained by simply disorder-averaging without invoking replica at all. See also~\cite{HH} for holographic study of quenched disorder using the memory function method.} We will use this criterion as a test on the machinery we develop in this paper to compute thermodynamic quantities.

\subsection{Power of holography}
It is in general not straightforward to compute disorder-averaged observables.  For example, to obtain the disorder-averaged grand potential density, we must compute 
\be
\le[\frac{\Omega}{V_{d-1}}\ri]_{\rm d.a.}\equiv\frac{1}{V_{d-1}}\int{\cal D}W P_V[W] \le\{ {\rm ln} \le( Z_{W({\bf x})} \ri)\ri\},
\ee
which is not equal to $\frac{1}{V_{d-1}}{\rm ln}\le\{\int{\cal D}W P_V[W] \le( Z_{W({\bf x})} \ri)\ri\}$.
Computing the logarithm of the partition function first and then disorder-averaging (not the other way around) is generally hard: dealing with this usually involves a handful of formal tricks, such as the replica trick and the cavity method. Looking through the holographic lens, our job is considerably simplified by the fact that the logarithm of the partition function is automatically computed by the gravitational action of the disordered geometry. This, together with the holographic geometrization of the functional flow, makes holography a computationally tractable playground for studying certain aspects of quenched disorder in a strongly correlated CFT.

\section{Intermezzo: Bulk Information}\label{rugged}
To systematically compute thermodynamic quantities and transport coefficients in this dirty holographic setup, we need to compute corrections to the geometry induced by the disordered boundary conditions.  In this section we compute these corrections perturbatively in the strength of the disorder in two steps.  First, we compute the leading bulk profile of the matter fields induced by their disordered boundary conditions.  We then use this profile to compute a self-averaged matter energy-momentum tensor in the bulk, and use the resulting homogenous energy-momentum tensor to self-consistently compute the leading backreaction to the bulk metric at $O(f_{\rm dis})$.  

To set the stage, recall that the clean CFT at finite temperature and zero chemical potential is described holographically by the following black brane geometry~\cite{WittenThermal}, here expressed in Euclidean time $\tau=+it$:
\bea
g_{MN}dx^M dx^N&=&+f(r)d\tau^2+\frac{dr^2}{f(r)}+\frac{r^2}{L^2}\le(\sum_{i=1}^{d-1}dx_i^2\ri)\\
{\rm with}\ \ \ f(r)&\equiv&\frac{r^2}{L^2}\le(1-\frac{r_+^d}{r^d}\ri).
\eea
The Euclidean time $\tau$ has a periodicity $\frac{4\pi}{d}\frac{L^2}{r_{+}}$ so as to make the geometry regular. We must now compute the matter field profile subject to disordered boundary conditions in this undistorted geometry.

\subsection{Matter profile}\label{probeprofile}
\subsubsection{Maxwell dirt on pure anti-de Sitter (AdS)}
The quenched-disordered boundary conditions induce a nontrivial profile for the bulk $U(1)$ gauge field. Let us begin with the pure AdS solution where $r_{+}=0$. To first order in $V({\bf x})$,
\be\label{probe}
A_M dx^M=\le[\int\frac{d^{d-1} {\bf k}}{(2\pi)^{d-1}}e^{i{\bf k}\cdot{\bf x}}V({\bf k})\le\{\pure\le(L{\bf k}; \frac{r}{L}\ri)\ri\}\ri](-id\tau).
\ee
The bulk to boundary Green function  $\pure\le({\bf {\tilde k}}; \rho\ri)$ is defined through
\bea\label{Maxwell}
\le[\partial_{\rho}^2+\frac{(d-1)}{\rho}\partial_{\rho}-\frac{{\bf {\tilde k}}^2}{\rho^4}\ri]\pure\le({\bf {\tilde k}}; \rho\ri)&=&0\\
{\rm with}\ \ \ \pure\le({\bf {\tilde k}}; \rho=\infty\ri)&=&1\\
{\rm and}\ \ \ \pure\le({\bf {\tilde k}}; \rho=0\ri)&=&0.
\eea
The first equation is just the probe Maxwell equation of motion, the second is the asymptotic boundary condition (\ref{UVBC}), and the last is the requirement of regularity at the Poincar\'{e} horizon. These can be exactly solved by
\be
\pure\le({\bf {\tilde k}}; \rho\ri)=\pure\le(\frac{{\bf {\tilde k}}}{\rho}\ri)=\le\{\frac{2^{2-\frac{d}{2}}}{\Gamma\le(\frac{d-2}{2}\ri)}\ri\} \le(\frac{|{\bf {\tilde k}}|}{\rho}\ri)^{\frac{d-2}{2}}K_{\frac{d-2}{2}}\le(\frac{|{\bf {\tilde k}}|}{\rho}\ri)
\ee
where $K_{\frac{d-2}{2}}(x)$ is the modified Bessel function of the second kind.

\subsubsection{Maxwell dirt on hot black brane}\label{shit}
For $r_+\ne0$, we have
\be\label{probe2}
A_M dx^M=\le[\int\frac{d^{d-1} {\bf k}}{(2\pi)^{d-1}}e^{i{\bf k}\cdot{\bf x}}V({\bf k})\le\{\BB\le(\frac{L^2{\bf k}}{r_+}; \frac{r}{r_+}\ri)\ri\}\ri](-id\tau)
\ee
with the bulk Green function  $\BB\le({\bf {\tilde k}}; \rho\ri)$ now satisfying
\bea\label{Maxwell2}
\le[\partial_{\rho}^2+\frac{(d-1)}{\rho}\partial_{\rho}-\frac{{\bf {\tilde k}}^2}{\rho^4\le(1-\frac{1}{\rho^d}\ri)}\ri]\BB\le({\bf {\tilde k}}; \rho\ri)&=&0\\
{\rm with}\ \ \ \BB\le({\bf {\tilde k}}; \rho=\infty\ri)&=&1\\
{\rm and}\ \ \ \BB\le({\bf {\tilde k}}; \rho=1\ri)&=&0.
\eea
The last equality is the requirement of regularity at the black brane horizon.\footnote{Irregular solutions behave near the horizon as $\sim\le[1+\frac{{\bf {\tilde k}}^2}{d}(\rho-1){\rm ln}(\rho-1)+O(\rho-1)\ri]$.} For our purposes, the crucial property of $\BB\le({\bf {\tilde k}}; \rho\ri)$, which will be key in showing the absence of inconsistent divergences in thermodynamic quantities below and which is derived in Appendix~\ref{Bessel}, is its high-momentum behavior near the boundary at $\rho\gg1$:
\be\label{high}
\partial_\rho\le\{{\rm ln}\le(\frac{\BB}{\pure}\ri)\ri\}=O\le(\frac{1}{\rho^{d-1}{\tilde {\bf k}}^2}\ri)\ \ \ {\rm for}\ \ \ 1\ll|{\bf {\tilde k}}|<\rho.
\ee

\subsection{First-order backreaction}\label{first-order}

\begin{figure}[t]
\includegraphics[scale=0.56,angle=0]{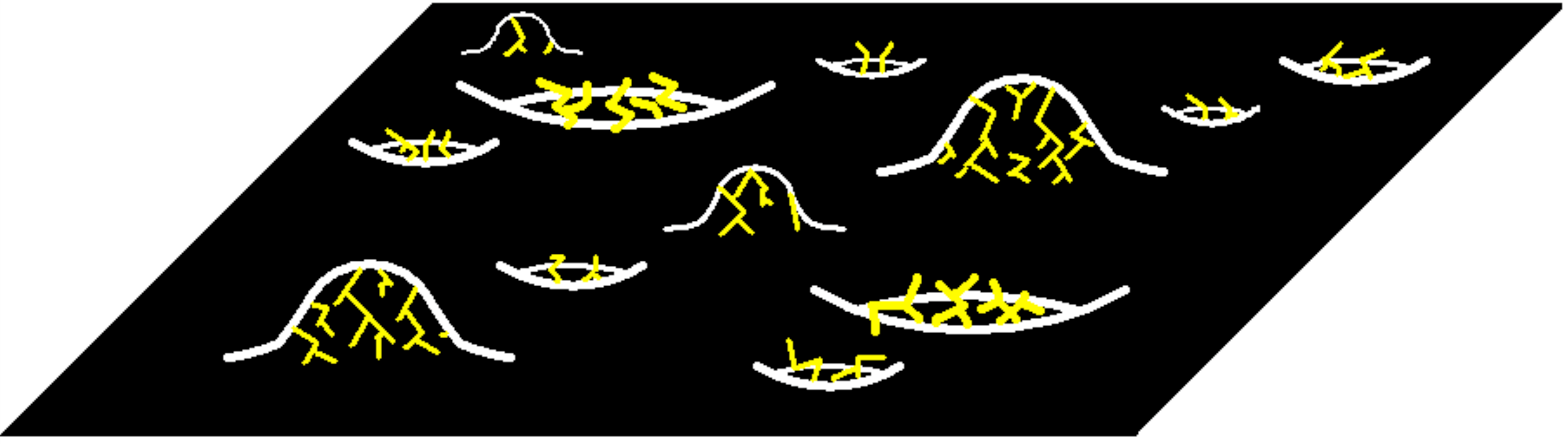}
\caption{An event horizon is distorted inhomogeneously by random electric field, giving rise to a rugged black brane.}
\label{1rugged}
\end{figure}

The nontrivial probe profile of the gauge field introduces nontrivial energy-momentum tensor $T_{MN}\equiv\frac{1}{g_{d+1}^2}\le[F_{MP}F_{N}^{\ P}-\frac{1}{4}F_{PQ}F^{PQ}g_{MN}\ri]$, seeding disorder nonlinearly into the right-hand side of the Einstein equation. Its backreaction then reshapes the black brane geometry into the rugged one (see Fig.~\ref{1rugged}). Though it is a complicated task to obtain even the leading correction to the geometry for the random potential $V({\bf x})$, the algebra simplifies at long distance. Namely, as we zoom out to long distance in the ${\bf x}$-direction, the quenched disorder self-averages and thus, in the bulk, an inhomogeneous energy-momentum tensor $T_{MN}$ self-averages into the homogeneous one $\le[T_{MN}\ri]_{\rm d.a.}$. Similarly, an inhomogeneous rugged geometry $g_{MN}$ self-averages into the homogeneous geometry $\le[g_{MN}\ri]_{\rm d.a.}$ at long distance.

To leading order, solving the Einstein equation with the homogeneous source $\le[T_{MN}\ri]_{\rm d.a.}$ yields the resulting homogeneous $\le[g_{MN}\ri]_{\rm d.a.}$. This is a straightforward, if stygian, exercise in gravitational perturbation theory, performed in Appendix~\ref{explicit} for the black brane geometry with the Gaussian disorder (\ref{distribution}). In this paper, we need an explicit expression only for a rugged pure AdS solution with $r_+=0$, which we record here:
\bea\label{ZeroBack}
\le[g_{MN}dx^M dx^N\ri]_{\rm d.a.}&=&+\le(\frac{r^2}{L^2}\ri)\le\{1+\epsilon_0\,\dfpure\!\le(\frac{r}{L}\ri)\ri\}d\tau^2+\frac{dr^2}{\le(\frac{r^2}{L^2}\ri)\le\{1+\epsilon_0\,\dfpure\!\le(\frac{r}{L}\ri)\ri\}}\nonumber\\
&&+\le(\frac{r^2}{L^2}\ri)\le\{1+\epsilon_0\,\dApure\!\le(\frac{r}{L}\ri)\ri\}\le(\sum_{i=1}^{d-1}dx_i^2\ri)
\eea
with $\epsilon_0\equiv f_{\rm dis}\times\le(\frac{G_{N}}{g_{d+1}^2}\frac{1}{L^2}\ri)\times\le(\frac{1}{L}\ri)^{d-3}$,
\bea
\dfpure(\rho)&=&\le(\frac{8\pi}{d-1}\ri)\le\{\frac{(2d-3)a_1+(2d-5)a_2}{(d-2)(2d-3)}\ri\}\rho^{d-3},\\
{\rm and}\ \ \ \dApure(\rho)&=&\le(\frac{8\pi}{d-1}\ri)\le\{\frac{-2a_2}{(d-2)}\ri\}\int_{\rho_2}^{\rho}d\rho'\rho'^{d-4}.
\eea
Here, $a_1\equiv\int\frac{d^{d-1} {\bf y}}{(2\pi)^{d-1}}{\bf y}^2\le\{\partial_y\pure(y)\ri\}^2$ and $a_2\equiv\int\frac{d^{d-1} {\bf y}}{(2\pi)^{d-1}}{\bf y}^2\le\{\pure(y)\ri\}^2$ are constants of order 1, related to each other by $(3d-5)a_1=(d+1)a_2$.\footnote{To derive this relation, use Eq.(\ref{miracle}) with $r_+=0$.} The integration constant $\rho_2$ will be appropriately chosen below.

\section{Functional Flows}\label{peek}

The real utility of the holographic approach is that it geometrizes functional flows.
Let us start with the pure AdS spacetime disordered by a random potential $V({\bf x})$ in the ultraviolet, characterized by a distribution $P_{V}[W({\bf x})]$. To see how the distribution runs as we change the energy scale, we can evolve $V({\bf x})$ from infinity down to some hypersurface at $r=r_{\star}$. This gives us $V_{\star}({\bf x})\equiv iA_{\tau}({\bf x},r_{\star})$, from which we can read off the corresponding distribution $P_{V_{\star}}[W({\bf x})]$ at energy scale $\sim \frac{r_{\star}}{L^2}$.

\begin{figure}[t]
\includegraphics[scale=0.50,angle=0]{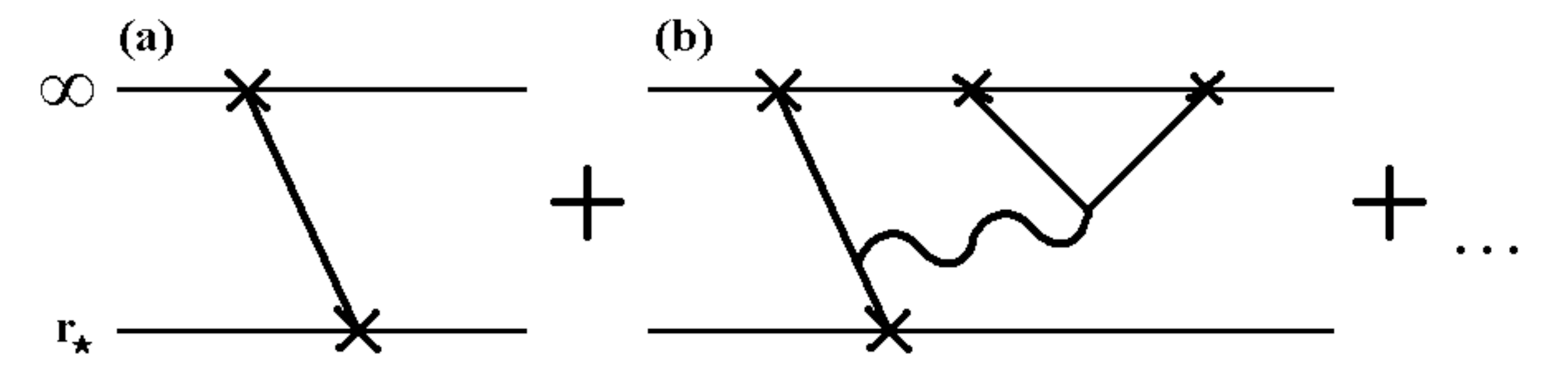}
\caption{The flow of disorder can be represented by Feynman-Witten diagrams. Solid lines represent probe propagations of $U(1)$ gauge fields governed by $G_0$ in the vacuum whereas a wavy line represents a graviton propagation.}
\label{2flow}
\end{figure}

\begin{figure}[t]
\includegraphics[scale=0.50,angle=0]{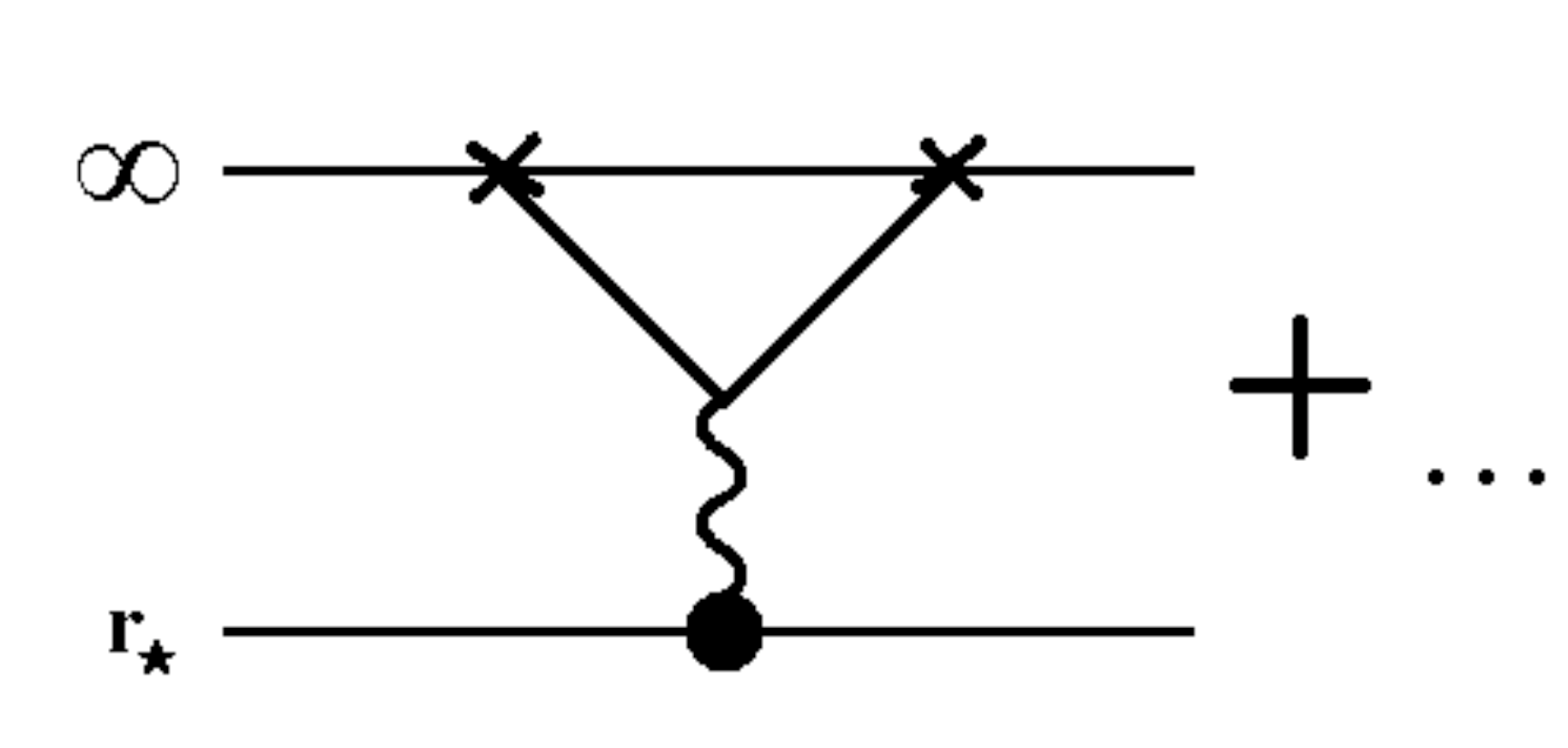}
\caption{Backreaction of disorder can also be represented by Feynman-Witten diagrams.}
\label{3back}
\end{figure}

Perturbatively, this process can be represented by Feynman-Witten diagrams (see Fig.~\ref{2flow}). In particular, holography provides an algorithmic way to keep track of functional flows.
We also need to keep track of how the disorder distorts the pure AdS geometry at a given energy scale. This evolution, too, can be represented by Feynman-Witten diagrams (see Fig.~\ref{3back}).

\subsection{Functional flow to first order in $f_{\rm dis}$}
Let us see how all this works at leading order in the strength of disorder. Formally flowing down to $r=r_{\star}$ convolves $V({\bf x})$ with the bulk-to-boundary Green function $\pure\le(\frac{L^2{\bf k}}{r_{\star}}\ri)$, yielding [see Fig.~\ref{2flow}(a)]
\be
 V_{\star}({\bf x})=\int\frac{d^{d-1} {\bf k}}{(2\pi)^{d-1}}e^{i{\bf k}\cdot{\bf x}}V({\bf k})\pure\le(\frac{L^2{\bf k}}{r_{\star}}\ri).
\ee
In particular, for the case where the random potential is governed by the Gaussian distribution (\ref{distribution}) at infinity, we obtain, at leading order in $f_{\rm dis}$, 
\be\label{stardistribution}
P_{V_{\star}}[W({\bf x})]=N_{\sharp, \star}e^{-\int \frac{d^{d-1}{\bf k}}{(2\pi)^{d-1}}\frac{1}{2f_{\star}({\bf k})}W({\bf k})W(-{\bf k})}\ \ \ {\rm with}\ \ \ f_{\star}({\bf k})=f_{\rm dis}\le\{\pure\le(\frac{L^2{\bf k}}{r_{\star}}\ri)\ri\}^2.
\ee

We have already analyzed the first-order backreaction to the pure AdS geometry in the previous section, but there is one subtlety: choice of the integration constant $\rho_2$.  It specifies the scale at which we define ``volume," which affects what we mean by ``density." We will stick to the choice $\rho_2=\infty$ for $d<2+1$ and $\rho_2=0$ for $d>2+1$ so that $\epsilon_0\dAfpure\le(\frac{r}{L}\ri)\propto f_{\rm dis}(\frac{r}{L^2})^{d-3}$, choosing conventions around the ultraviolet and infrared stable fixed points, respectively. Similar remarks apply for ``time" and ``temperature," as usual.

Note that, for $d>2+1$, the disorder is irrelevant and thus starts to plague the ultraviolet geometry at large $r$. Thus we need to carefully renormalize disorder-averaged observables as we take the cutoff scale $r_{\cut}$ to infinity for this class of deformations.
For $d<2+1$, temperature provides an infrared cutoff scale, shielding us from infrared catastrophes.
For $d=2+1$, while $\epsilon_0\dfpure\le(\frac{r}{L}\ri)\propto f_{\rm dis}$, we find a logarithmic running in $\le[g_{ii}\ri]_{\rm d.a.}$: $\epsilon_0\dApure\le(\frac{r}{L}\ri)\propto f_{\rm dis}{\rm ln}(\frac{r}{\rho_2 L})$. Whether this destroys the infrared geometry or the ultraviolet geometry depends on whether the deformation is marginally relevant or irrelevant. We will come back to this issue in the sequel~\cite{2ndAY} by going one order higher in $f_{\rm dis}$.

\section{Holographic Functional Renormalization}\label{fRG}
To compute disorder-averaged thermodynamic quantities at temperature $T$, we need to regulate them by cutting off the rugged black brane geometry at $r=r_{\cut}$, then specify Dirichlet boundary conditions for the bulk fields -- including the disorder functional -- at the cutoff surface, evaluate the on-shell action, subtract temperature-independent divergences, and finally take the limit $r_{\cut}\rightarrow\infty$.
Specifically, we propose a following recipe:\footnote{For simplicity, we employ a background-subtraction scheme. A more systematic treatment as in \cite{kostas} would be interesting to pursue.}
\begin{enumerate}

\item At $r=r_{\cut}$, fix the Dirichlet boundary conditions for the rugged black brane solution to be same as those of the rugged pure AdS solution with thermal-time periodicity $\tau_{\rm period}=\frac{1}{T}$ and volume $V_{d-1}$.

\item Compute the regulated on-shell Euclidean action, which we identify as $\frac{\Omega(T)}{T}$ via the holographic dictionary, for the rugged black brane solution. Renormalize away temperature-independent divergences by subtracting the $r=r_{\cut}$ surface contribution from the on-shell action for the rugged pure AdS solution. Take $r_{\cut}\rightarrow\infty$.

\end{enumerate}

\subsection{Scheme in action to first order in $f_{\rm dis}$}
In the special case of quenched electric disorder governed by a Gaussian distribution (\ref{distribution}) in the ultraviolet, this scheme works out as follows:

\begin{enumerate}

\item We first find the rugged black brane solution whose $V_{\cut}({\bf x})\equiv \le(1-\frac{r_+^d}{r_{\cut}^d}\ri)^{-\frac{1}{2}}iA_{\tau}({\bf x},r_{\cut})$ is distributed according to [c.f. Eq.(\ref{stardistribution})]\footnote{Note that we need to compare the random $U(1)$ gauge field potential on the black brane geometry and the one on the pure AdS geometry in a properly rescaled time coordinate at $r=r_{\cut}$, resulting in the extra factor of $\le(1-\frac{r_+^d}{r_{\cut}^d}\ri)^{-\frac{1}{2}}$ above.}
\be\label{cutdistribution}
P_{V_{\cut}}[W({\bf x})]=N_{\sharp, \cut}e^{-\int \frac{d^{d-1}{\bf k}}{(2\pi)^{d-1}}\frac{1}{2f_{\cut}({\bf k})}W({\bf k})W(-{\bf k})}\ \ \ {\rm with}\ \ \ f_{\cut}({\bf k})=f_{\rm dis}\le\{\pure\le(\frac{L^2{\bf k}}{r_{\cut}}\ri)\ri\}^2
\ee
and whose Euclidean time periodicity $\tau_{\rm period}$ and volume $V_{d-1}$ are set by
\bea
\tau_{\rm period}\sqrt{\le[g_{\tau\tau}(r_{\cut})\ri]_{\rm d.a.}}&=&\le(\frac{r_{\cut}}{L}\ri)\le\{1+ \frac{1}{2}\epsilon_0 \dfpure\le(\frac{r_{\cut}}{L}\ri)\ri\}\frac{1}{T}\\
{\rm and}\ \ \ \int d^{d-1}{\bf x}\le(\sqrt{\le[g_{ii}(r_{\cut})\ri]_{\rm d.a.}}\ri)^{d-1}&=&\le(\frac{r_{\cut}}{L}\ri)^{d-1}\le\{1+\frac{(d-1)}{2}\epsilon_0 \dApure\le(\frac{r_{\cut}}{L}\ri)\ri\}V_{d-1}.
\eea

\item We then compute the on-shell action for the rugged black brane solution obtained above, subtract temperature-independent divergences, and take the limit $r_{\cut}\rightarrow\infty$.  The Maxwell action contributes a temperature-independent divergence for the disorder-averaged grand potential density of the form\footnote{Here, we are working with grandcanonical ensemble at zero average chemical potential. To work with canonical ensemble requires us to add an appropriate boundary term, changing Dirichlet boundary condition to Neumann boundary condition.}
\be\label{Maxwellshit}
-\frac{1}{2g_{d+1}^2}f_{\rm dis}\int\frac{d^{d-1} {\bf k}}{(2\pi)^{d-1}}\le\{\pure\le(\frac{L^2{\bf k}}{r_{\cut}}\ri)\ri\}^2\le(\frac{r_{\cut}}{L}\ri)^{d-1}\le[\frac{\partial_r \pure}{\pure}\ri]\Big|_{r=r_{\cut}}
\ee
while the contribution from Einstein action evaluates to
\be\label{Einsteinshit}
-\frac{(d-1)L^{d-1}}{8\pi G_N}\le(\frac{r_{\cut}}{L^2}\ri)^d \times\le[1+\frac{\epsilon_0}{2}\dfpure\le(\frac{r_{\cut}}{L}\ri)+\frac{(d-1)\epsilon_0}{2}\dApure\le(\frac{r_{\cut}}{L}\ri)\ri].
\ee
An easy way to get the latter is to note that, with the standard Gibbons-Hawking surface term, the only correction at $O(f_{\rm dis})$ with respect to the undistorted pure AdS geometry comes from the change in the Dirichlet boundary condition at $r=r_{\cut}$.

\end{enumerate}
On the field theory side, what the functional (\ref{cutdistribution}) at $r=r_{\cut}$ succinctly codifies is a complicated distribution entailing the cornucopia of multi-trace random deformations generated by integrating out the geometry from $r=\infty$ to $r=r_{\cut}$ \cite{FLR,HP}.
Note that, due to the exponential decay of $\pure(y)=\le\{\frac{2^{2-\frac{d}{2}}}{\Gamma\le(\frac{d-2}{2}\ri)}\ri\} y^{\frac{d-2}{2}}K_{\frac{d-2}{2}}\le(y\ri)$ at large $y$, the momentum is cut off above $|{\bf k}|\sim \frac{r_{\cut}}{L^2}$, conforming with the standard holographic intuition.\footnote{Had there been no such effective cutoff, we would have suffered from infinite backreaction to the geometry.}

\section{Thermodynamics}\label{thermodynamics}
We are now ready to calculate the disorder-averaged grand potential density $\le[\frac{\Omega(T)}{V_{d-1}}\ri]_{\rm d.a.}$ to first order in $f_{\rm dis}$, following the scheme developed above.
The eventual satisfaction of the Harris criterion provides a nontrivial check on the machinery developed herein.

\subsection{Contribution from Maxwell}
Given the distribution (\ref{cutdistribution}), the $O(f_{\rm dis})$ contribution from the Maxwell action to the disorder-averaged grand potential density becomes
\be
-\frac{1}{2g_{d+1}^2}f_{\rm dis}\int\frac{d^{d-1} {\bf k}}{(2\pi)^{d-1}}\le\{\pure\le(\frac{L^2{\bf k}}{r_{\cut}}\ri)\ri\}^2\le(\frac{r_{\cut}}{L}\ri)^{d-1}\le(1-\frac{r_+^d}{r_{\cut}^d}\ri)^{-\frac{1}{2}}\le[\frac{\partial_r \BB}{\BB}\ri]\Big|_{r=r_{\cut}}.
\ee
Subtracting the temperature-independent divergence (\ref{Maxwellshit}), we get 
\bea
\le[\frac{\Omega_{\rm Maxwell}}{V_{d-1}}\ri]_{\rm d.a.}&=&-\frac{\nmax^2}{2}f_{\rm dis}\le(\frac{4\pi T}{d}\ri)^{2d-3}\nonumber\\
&&\times\int \frac{d^{d-1}{\tilde {\bf k}}}{(2\pi)^{d-1}}\le[\rho^{d-1}\pure^2\partial_\rho\le\{{\rm ln}\le(\frac{\BB}{\pure}\ri)\ri\}-\frac{1}{2\rho}\pure^2\partial_\rho\le\{{\rm ln}\le(\BB\ri)\ri\}\ri]\Big|_{\rho=\rho_{\cut}}.
\eea
where $\rho_{\cut}\equiv\frac{r_{\cut}}{r_+}$.

At low momentum, the integrand in the last parenthesis behaves well: for example, $\le[\rho^{d-1}\pure^2\partial_\rho\le\{{\rm ln}\le(\frac{\BB}{\pure}\ri)\ri\}\ri]\Big|_{\rho=\rho_{\cut}}$ approaches a finite number, $(d-2)$, since $\BB\le({\bf {\tilde k}}=0; \rho\ri)=1-\frac{1}{\rho^{d-2}}$. However, for $d\geq2+1$, contributions from high momentum modes with ${\tilde {\bf k}}\sim \rho_{\cut}$ give rise to severe divergences: to see how the first term diverges, let us differentiate it with respect to $\rho_{\cut}$:
\bea
&&\partial_{\rho_{\cut}}\int \frac{d^{d-1}{\tilde {\bf k}}}{(2\pi)^{d-1}}\le[\rho^{d-1}\pure^2\partial_\rho\le\{{\rm ln}\le(\frac{\BB}{\pure}\ri)\ri\}\ri]\Big|_{\rho=\rho_{\cut}}\nonumber\\
&=&\int \frac{d^{d-1}{\tilde {\bf k}}}{(2\pi)^{d-1}}\le[\rho^{d-1}G_0^2\le\{\frac{{\tilde {\bf k}}^2}{\rho^4(\rho^d-1)}-\le[\partial_\rho\le\{{\rm ln}\le(\frac{\BB}{\pure}\ri)\ri\}\ri]^2\ri\}\ri]\Big|_{\rho=\rho_{\cut}}\nonumber\\
&=&a_2\rho_{\cut}^{d-4}+O\le(\frac{1}{\rho_{\cut}^4}\ri)
\eea
where we used the property (\ref{high}) in the last step. Similarly, the second term in the parenthesis can be massaged into the form
\bea
&&\int \frac{d^{d-1}{\tilde {\bf k}}}{(2\pi)^{d-1}}\le[-\frac{1}{2\rho}\pure^2\partial_\rho\le\{{\rm ln}\le(\BB\ri)\ri\}\ri]\Big|_{\rho=\rho_{\cut}}\nonumber\\
&=&-\int \frac{d^{d-1}{\tilde {\bf k}}}{(2\pi)^{d-1}}\le[\frac{1}{2\rho^d}\le(\rho^{d-1}\pure\partial_\rho\pure\ri)\ri]\Big|_{\rho=\rho_{\cut}}+O\le(\frac{1}{\rho_{\cut}^3}\ri)\nonumber\\
&=&-\int \frac{d^{d-1}{\tilde {\bf k}}}{(2\pi)^{d-1}}\le[\frac{1}{2\rho^d}\int_0^{\rho}d\rho'\rho'^{d-1}\le\{(\partial_\rho\pure)^2+\frac{{\tilde {\bf k}}^2}{\rho'^4}G_0^2\ri\}\ri]\Big|_{\rho=\rho_{\cut}}+O\le(\frac{1}{\rho_{\cut}^3}\ri)\nonumber\\
&=&-\frac{(a_1+a_2)}{2(2d-3)}\rho_{\cut}^{d-3}+O\le(\frac{1}{\rho_{\cut}^3}\ri).
\eea
All in all, we end up with\footnote{For $d=2+1$, replace $\frac{1}{(d-3)}\rho_{\cut}^{d-3}$ by ${\rm ln}(\rho_{\cut})$.}
\be
\le[\frac{\Omega_{\rm Maxwell}}{V_{d-1}}\ri]_{\rm d.a.}=-\frac{\nmax^2}{2}f_{\rm dis}\le(\frac{4\pi T}{d}\ri)^{2d-3} \le[({\rm finite\ piece})+\le\{-\frac{(a_1+a_2)}{2(2d-3)}+\frac{a_2}{(d-3)}\ri\}\rho_{\cut}^{d-3}\ri].
\ee
The divergent coefficient multiplying a temperature-dependent term looks horrifying at first.
However, this divergence is precisely cancelled by a matching term in the Einstein action.

\subsection{Contribution from Einstein}
To evaluate the on-shell Einstein action for the rugged black brane geometry, we can employ the same trick we used for the rugged pure AdS geometry. The crucial step is to use the undistorted black brane geometry on-shell action
\be
-\frac{(d-1)L^{d-1}}{8\pi G_N}\le(\frac{r_{\cut}}{L^2}\ri)^d V'_{d-1}\times\le\{1-\frac{(d-2)}{2(d-1)}\frac{r_+^d}{r_{\cut}^d}\ri\}\times\frac{4\pi}{d}\frac{L^2}{r_+}
\ee
with the modified volume $V'_{d-1}=\le\{1+\frac{(d-1)}{2}\epsilon_0 \dApure\le(\frac{r_{\cut}}{L}\ri)\ri\}V_{d-1}$ and the horizon size $r_+$ set by
\be
\frac{4\pi}{d}\frac{L^2}{r_+}\sqrt{1-\frac{r_+^d}{r_{\cut}^d}}=\le\{1+ \frac{1}{2}\epsilon_0 \dfpure\le(\frac{r_{\cut}}{L}\ri)\ri\}\frac{1}{T}.
\ee
Subtracting the $T$-independent divergence (\ref{Einsteinshit}) and taking the limit $r_{\cut}\rightarrow\infty$, we obtain
\bea
\le[\frac{\Omega_{\rm Einstein}}{V_{d-1}}\ri]_{\rm d.a.}&=&-\le\{\frac{1}{4d}\le(\frac{4\pi}{d}\ri)^{d-1}\ri\}\nein^2 T^d\nonumber\\
&&-\frac{\nmax^2}{2}f_{\rm dis}\le(\frac{4\pi T}{d}\ri)^{2d-3} \le[\le(\frac{d-1}{16\pi}\ri)\le(\frac{L}{r_+}\ri)^{d-3}\le\{-\dfpure\le(\frac{r_{\cut}}{L}\ri)+\dApure\le(\frac{r_{\cut}}{L}\ri)\ri\}\ri]\nonumber\\
&=&-\le\{\frac{1}{4d}\le(\frac{4\pi}{d}\ri)^{d-1}\ri\}\nein^2 T^d\\
&&-\frac{\nmax^2}{2}f_{\rm dis}\le(\frac{4\pi T}{d}\ri)^{2d-3} \le[\le\{-\frac{(2d-3)a_1+(2d-5)a_2}{2(d-2)(2d-3)}-\frac{a_2}{(d-2)(d-3)}\ri\}\rho_{\cut}^{d-3}\ri].\nonumber
\eea
Upon using the identity $(3d-5)a_1=(d+1)a_2$, we see that the divergent term here precisely cancels the divergence we found in the Maxwell action, $\le[\frac{\Omega_{\rm Maxwell}}{V_{d-1}}\ri]_{\rm d.a.}$.

\subsection{Grand potential density: satisfaction of Harris}
With the naive divergences consistently cancelled out, we finally obtain
\be
\le[\frac{\Omega(T)}{V_{d-1}}\ri]_{\rm d.a.}= c_0 \nein^2 T^d +c_1 \nmax^2  f_{\rm dis} T^{2d-3}+O(f_{\rm dis}^2)
\ee
where $c_0=-\frac{1}{4d}\le(\frac{4\pi}{d}\ri)^{d-1}$ and $c_1$ are constants of order 1.
This formula happily accords with the Harris criterion.

\section{Impure Thoughts}\label{conclusion}
In this paper, we have studied strongly correlated CFTs holographically dual to classical Einstein-Maxwell theory in the presence of the quenched electric disorder. In particular, we developed a holographic functional renormalization scheme and, for Gaussian disorder, computed the disorder-averaged grand potential density to first order in the strength of the disorder, $f_{\rm dis}$. The result accords with the Harris criterion, which adds confidence to the validity of our scheme. Namely, the quenched electric disorder dominates low-temperature thermodynamics for $d<2+1$, whereas its effects essentially disappear at low temperature for $d>2+1$. A particularly interesting case was $d=2+1$ for which the quenched disorder was found to be marginal. 
In a forthcoming paper~\cite{2ndAY}, we will investigate whether the quenched electric disorder is marginally relevant or irrelevant for $d=2+1$.

These results indicate that holographic perturbation theory in $f_{\rm dis}$ becomes unreliable at sufficiently low temperature for $d<2+1$ (and possibly $d=2+1$). Naturally, we would expect dramatic phenomena to emerge precisely when such perturbative analysis breaks down and the bulk geometry is significantly distorted.
Thus we wish to embark on the journey beyond the perturbative regime, looking for transitions/crossovers lurking behind.

Several powerful nonperturbative techniques have been developed in the study of spin glasses~\cite{Spinglass1,Spinglass2}.
For example, these techniques enable us to see the glass transition in the Sherrington-Kirkpatrick model for which classical mean field theory is valid.
In particular, the replica method has been an extremely useful tool to analyze disordered systems in considerable generality.
We wish to bring these techniques to bear on our particular problem at hand, which admits a dual representation in terms of classical gravitational theory.\footnote{This is nontrivial when $\frac{\nmax}{\nein}\sim1$. The simple application of the replica method to the problem at hand produces double-trace operators which scale as $\nein^4$, whereas a simple holographic prescription is known for double-trace operators only when they scale as $\nein^2$ or less~\cite{MultiWitten}.}

One natural scenario motivated by analogy to thermodynamic behavior of the Sherrington-Kirkpatrick model would be the following: for $d<2+1$, as we decrease the temperature, the effective strength of the disorder $f_{\rm dis}T^{d-3}$ grows, with perturbation theory breaking down when $f_{\rm dis}T^{d-3}\sim 1$. At this point, the system may enter into a glassy phase where we have not just one but many metastable solutions for a given asymptotic boundary condition $V({\bf x})$. It would also be exciting to see any connection to percolating picture of variable-range hopping for weakly correlated systems~\cite{Percolation}, for example a fragmentation of the black brane horizon.

\begin{acknowledgments}
We thank Oliver DeWolfe, Sean Hartnoll, Shamit Kachru, Steven Kivelson, Hong Liu, Joseph Maciejko, Michael Mulligan, Shinsei Ryu, Stephen Shenker, and Eva Silverstein for very useful discussions.
S.Y. thanks Hong Liu and the Center for Theoretical Physics at Massachusetts Institute of Technology for their generous hospitality when this work was initiated, A.A. thanks the Stanford Institute for Theoretical Physics for hospitality while this work was being completed, and we both thank the organizers and participants of the Aspen 2011 winter conference ``Strongly Correlated Systems and Gauge/Gravity Duality" for providing a stimulating environment to polish up the current paper.
The research of A.A. is supported by the DOE under contract \#DE-FC02-94ER40818.
S.Y. is supported by the Stanford Institute for Theoretical Physics and NSF Grant No. 0756174.
\end{acknowledgments}

\appendix

\section{Asymptotic expansion with Bessel envelope}
\label{Bessel}
In this appendix, we analyze solutions to the probe Maxwell equation (\ref{Maxwell2}) in the black brane geometry in the high momentum limit.\footnote{For (and only for) $d=2+1$, the result in this appendix can be reproduced by simpler WKB asymptotic expansion.}
To start with, we focus on the region very close to the boundary by going to the coordinate $y=\frac{|{\tilde {\bf k}}|}{\rho}$:
\be\label{newMaxwell}
\le[\partial_y^2+\frac{3-d}{y}\partial_y-\frac{1}{1-\frac{y^d}{|{\tilde {\bf k}}|^d}}\ri]\phi=0.
\ee
If we neglect $\frac{y^d}{|{\tilde {\bf k}}|^d}$ for a moment, the equation is exactly solved by
\be
\pure(y)\equiv\le\{\frac{2^{2-\frac{d}{2}}}{\Gamma\le(\frac{d-2}{2}\ri)}\ri\} y^{\frac{d-2}{2}} K_{\frac{d-2}{2}}(y).
\ee
For $y\ll1$, it approaches $1$, whereas for $y\gg1$, it exponentially decays as $y^{\frac{d-3}{2}}e^{-y}$.

We will try to find a positive shrinking solution $\phi_{\rm shrinking}$ for Eq.(\ref{newMaxwell}) with $\phi_{\rm shrinking}({\tilde {\bf k}}; y=0)=1$ which, for large $|{\tilde {\bf k}}|$, rapidly decreases as we move away from the boundary at $y=0$. There also exists a positive growing solution $\phi_{\rm growing}$, say with the near boundary behavior $\phi_{\rm growing}({\tilde {\bf k}}; y)=0+1\times y^{d-2}+...$, which rapidly increases away from the boundary for large $|{\tilde {\bf k}}|$. The regular probe solution $\BB$ with $\BB\le({\bf {\tilde k}}; \rho=\infty\ri)=1$ and $\BB\le({\bf {\tilde k}}; \rho=1\ri)=0$ is a linear combination of the two, but the coefficient in front of $\phi_{\rm growing}$ must be exponentially small for large $|{\tilde {\bf k}}|$ so as to satisfy the boundary condition at the horizon $\rho=1$. Therefore, up to exponentially suppressed contribution, the high-momentum behavior of $\BB$ is entirely governed by $\phi_{\rm shrinking}$.

We now generate an asymptotic series for $\phi_{\rm shrinking}$ by using $\pure$ as an envelope:
\bea
\phi_{\rm shrinking}({\tilde {\bf k}}; y)&=&\pure(y)\le[\sum_{n=0}^{\infty}\frac{1}{|{\tilde {\bf k}}|^{nd}}\psi_n(y)\ri]\\
{\rm with}\ \ \ \psi_0(y)&=&1.
\eea
Plugging this into the probe equation yields a recursive relation
\bea
\psi_n(y)&=&d_n \le\{(d-2)\int_0^ydy'\frac{y'^{d-3}}{\pure^2(y')}\ri\}\nonumber\\
&&+\int_0^ydy'\frac{y'^{d-3}}{\pure^2(y')}\int_0^{y'}dy''\pure^2(y'')y''^3\le\{\sum_{l=0}^{n-1}y''^{(n-1-l)d}\psi_l(y'')\ri\}.
\eea
Here the $d_n$'s are constants specifying the leading normalizable piece at each order in $\frac{1}{|{\tilde {\bf k}}|^d}$ expansion. For a generic choice of $d_n$'s, the corresponding solution grows exponentially for large $y$. Since we are looking for a shrinking solution, we will make a special choice of $d_n$'s to tame such a rapid growth. Namely, we recursively choose
\be
d^{\rm shrinking}_n=-\frac{1}{(d-2)}\int_0^{\infty}dy\pure^2(y)y^3\le\{\sum_{l=0}^{n-1}y^{(n-1-l)d}\psi_l(y)\ri\}.
\ee

With this particular choice, we can inductively show that $\psi_n(y)$ grows only as $y^{(d+1)n}$ for large $y$ as opposed to generic exponential growth. In particular, the series provides a nice asymptotic expansion as long as $y\ll|{\tilde {\bf k}}|^{\frac{d}{d+1}}$. With the envelope, it follows that this special solution is in fact shrinking exponentially whereas generic solutions are exponentially growing. This shrinking solution (and hence $\BB$) has the property advertised in Sec.\ref{shit}:
\be
\partial_\rho\le\{{\rm ln}\le(\frac{\phi_{\rm shrinking}}{\pure}\ri)\ri\}=O\le(\frac{1}{\rho^{d-1}{\tilde {\bf k}}^2}\ri)\ \ \ {\rm for}\ \ \ 1\ll|{\bf {\tilde k}}|<\rho.
\ee

\section{First-order backreaction with the Gaussian distribution}
\label{explicit}

To first order in $f_{\rm dis}$, defining $\epsilon_+\equiv f_{\rm dis}\times\le(\frac{G_{N}}{g_{d+1}^2}\frac{1}{L^2}\ri)\times\le(\frac{r_+}{L^2}\ri)^{d-3}$ and
\bea
s_1(\rho)&\equiv&\le(\frac{8\pi}{d-1}\ri)\int\frac{d^{d-1} {\tilde {\bf k}}}{(2\pi)^{d-1}}|\partial_{\rho}\BB|^2,\\
s_2(\rho)&\equiv&\le(\frac{8\pi}{d-1}\ri)\int\frac{d^{d-1} {\tilde {\bf k}}}{(2\pi)^{d-1}}\le[\frac{{\bf {\tilde k}}^2}{\rho^4\le(1-\frac{1}{\rho^d}\ri)}|\BB|^2\ri],
\eea
we have
\bea
8\pi G_{N}\le[T_{\tau\tau}\ri]_{\rm d.a.}&=&\frac{(d-1)\epsilon_+}{2L^2}\times f(r)\times \le[-s_1(\rho)-s_2(\rho)\ri],\\
8\pi G_{N}\le[T_{rr}\ri]_{\rm d.a.}&=&\frac{(d-1)\epsilon_+}{2L^2}\times\frac{1}{f(r)}\times \le[-s_1(\rho)+s_2(\rho)\ri],\\
8\pi G_{N}\le[T_{ij}\ri]_{\rm d.a.}&=&\frac{(d-1)\epsilon_+}{2L^2}\times\frac{r^2}{L^2}\delta_{ij}\times
\le[s_1(\rho)+\le(\frac{d-3}{d-1}\ri)s_2(\rho)\ri],
\eea
with all the other components vanishing.
As the self-averaged energy-momentum tensor is homogeneous, without loss of generality, we make the following ansatz for the self-averaged geometry:
\bea\label{selfaveragedgeometry}
\le[g_{MN}dx^M dx^N\ri]_{\rm d.a.}&=&+f(r)\le\{1+\epsilon_+\,\df\!\le(\frac{r}{r_+}\ri)\ri\}d\tau^2+\frac{dr^2}{f(r)\le\{1+\epsilon_+\,\df\!\le(\frac{r}{r_+}\ri)\ri\}}\nonumber\\
&&+\frac{r^2}{L^2}\le\{1+\epsilon_+\,\dA\!\le(\frac{r}{r_+}\ri)\ri\}\le(\sum_{i=1}^{d-1}dx_i^2\ri).
\eea
We made a coordinate choice in $r$ to set $\le[g_{\tau\tau}\ri]_{\rm d.a.}=\le[g^{rr}\ri]_{\rm d.a.}$. Then, plugging it into Einstein equation and expanding to first order in $\epsilon_+$, we get ordinary differential equations for $\df(\rho)$ and $\dA(\rho)$. After lengthy manipulations, we arrive at following regular solutions:
\bea
\df(\rho)&=&\frac{1}{\rho^d-1}\int_1^{\rho}d\rho'\le[\le\{2(d-1)\rho'^{d-2}-\le(\frac{d-2}{\rho'^2}\ri)\ri\}\le\{\int_{\rho_0}^{\rho'}d\rho''s_3(\rho'')\ri\}\ri]\\ \label{dAC}
{\rm and}\ \ \ \dA(\rho)&=&-\int_{\rho_2}^{\rho}d\rho'\le[\le\{\frac{2}{\rho'^2}\int_{\rho_0}^{\rho'}d\rho''s_3(\rho'')\ri\}+\frac{\le\{s_1(\rho')-s_2(\rho')\ri\}}{\rho'\le\{(d-1)-\le(\frac{d-2}{2}\ri)\frac{1}{\rho'^d}\ri\}}\ri]\\
{\rm with}\ \ \ s_3(\rho)&\equiv&\frac{\rho^d\le\{2(d-1)(2d-3)\rho^d-(d-2)(d-3)\ri\}s_1(\rho)}{\le\{2(d-1)\rho^d-(d-2)\ri\}^2}\nonumber\\
&&+\frac{\rho^d\le\{2(d-1)(2d-5)\rho^d-(d-2)(3d-5)\ri\}s_2(\rho)}{\le\{2(d-1)\rho^d-(d-2)\ri\}^2}.
\eea
Here $\rho_0$ is a free parameter related to a constant coordinate shift in $r$ and $\rho_2$ is another integration constant. We can simplify the expressions further by using the identity
\be\label{miracle}
s_3(\rho)=\frac{s_2(\rho)}{\le(1-\frac{1}{\rho^d}\ri)}-\frac{d}{d\rho}\le[\frac{\rho\le\{s_1(\rho)-s_2(\rho)\ri\}}{\le\{2(d-1)-\frac{(d-2)}{\rho^d}\ri\}}\ri],
\ee
which follows from Eq.(\ref{Maxwell2}) for $\BB(\rho)$. Taking the limit $r_+\rightarrow0$ of this solution yields the rugged pure AdS solution (\ref{ZeroBack}).

\end{document}